\documentclass{elsarticle}
\usepackage{amssymb}
\usepackage{graphicx}
\usepackage{amsmath}
\usepackage{color}
\usepackage{hyperref}

\begin{document}
\title{A weak entanglement approximation for nuclear structure: a progress report}

\author{Calvin W. Johnson}
\affiliation{Department of Physics, San Diego State University, 5500 Campanile Drive, San Diego, CA 92182-1233, United States}

\author{Oliver C. Gorton}
\affiliation{Lawrence Livermore National Laboratory, P.O. Box 808, L-414, Livermore, California 94551, United States}

\date{\today}
\begin{abstract}
We report on a recently proposed approach, inspired by quantum information
theory,  for calculating low-energy nuclear structure  in the framework of the
configuration-interaction shell-model.  Empirical evidence has demonstrated
that the many-proton and many-neutron partitions of nuclear
configuration-interaction wave functions are weakly entangled, especially away
from $N=Z$. This has been developed into a practical methodology, the Proton
And Neutron Approximate Shell-model (PANASh). We review the basic ideas and
present  recent results. We also discuss some technical developments in
calculations.
\end{abstract}
\maketitle

\section{Introduction}

One venerable approach to nuclear structure is the configuration-interaction
method using a basis of shell-model configurations~\cite{br88,ca05}, though it
is by no means the only one.  One expands the wave function in a basis,
\begin{equation}
| \Psi \rangle = \sum_{\alpha} c_\alpha | \alpha \rangle,
\end{equation}
and then finds the stationary states by solving a matrix eigenvalue problem. 

Now one has to choose the basis, $\{ | \alpha \rangle \}$.  One can choose
complex basis states that embody many correlations, but in that case
constructing the states and computing the matrix elements of the Hamiltonian
can be very time consuming. Alternately, one can choose very simple basis
states, for example Slater determinants (or, more properly, the occupation
representation of Slater determinants using second quantization), for which
there are fast methods to compute Hamiltonian matrix elements
on-the-fly~\cite{BIGSTICK}, but then the number of basis states need to build
up physical correlations can be very large.  Because the nuclear Hamiltonian is
rotationally invariant,  many nuclear configuration-interaction codes work with
bases with fixed $J_z$ or $M$, called the $M$-scheme. The current largest
$M$-scheme calculations utilize around $3 \times 10^{10}$ basis
states~\cite{mccoy2024intruder}. Nonetheless many systems of interest have
dimensions far beyond this limit. 

While there are many possible truncation schemes, a recent
approach~\cite{PhysRevC.110.034305} builds upon ideas from quantum information
theory~\cite{johnson2023proton}.  By breaking the problem into two pieces,
solving independently, and then combining, one finds an effective and practical
truncation that could extend the reach of the configuration-interaction
shell-model approach. In Section~\ref{WEA} we introduce the motivation and
formalism for a ``weak entanglement approximation,'' followed by some sample
results in Section~\ref{results}.  In \ref{technical}, we discuss some
technical challenges and how they have been recently mitigated.

\begin{figure}
    \centering
    \includegraphics[width=0.6\textwidth]{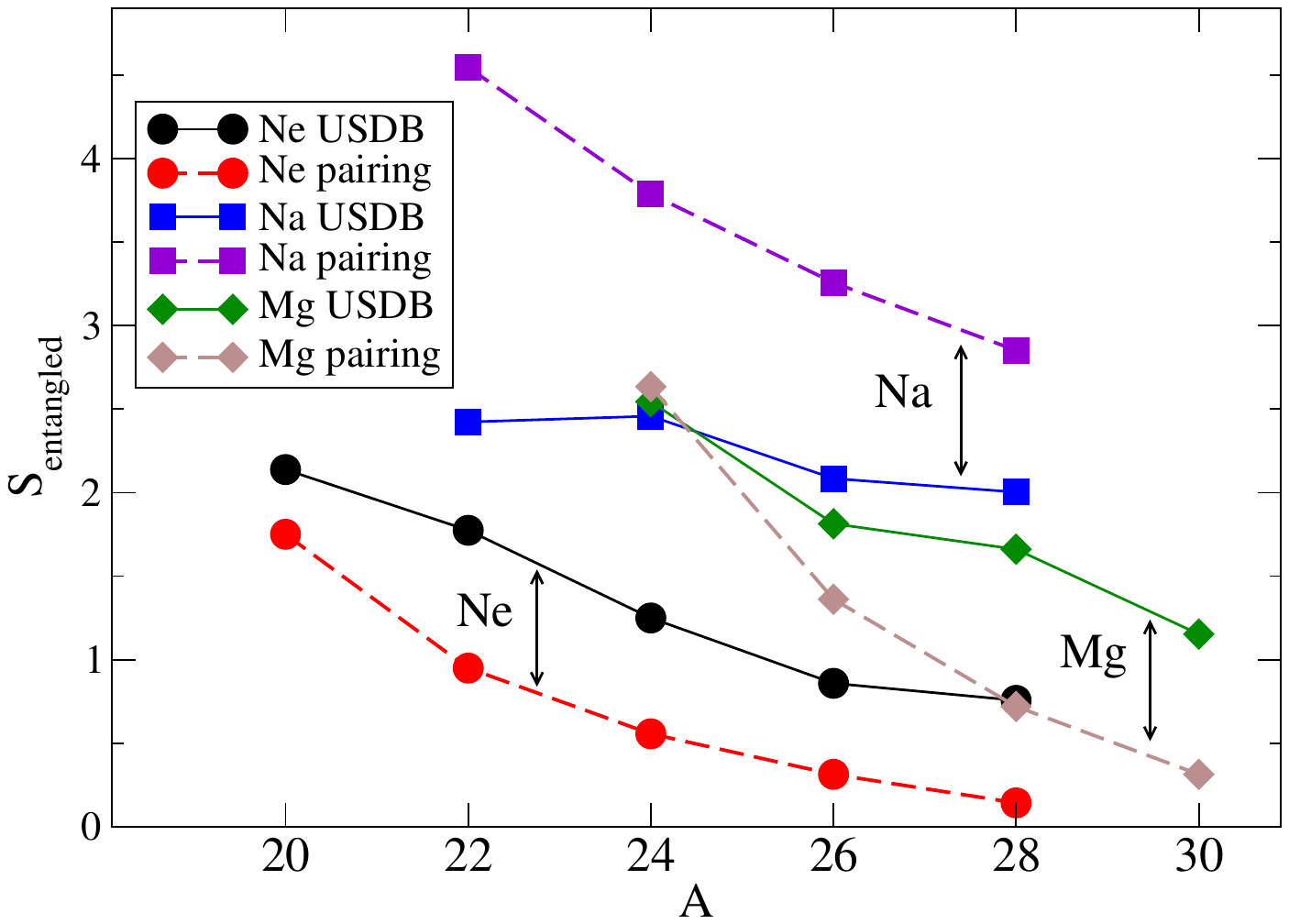}
    \caption{Entropy of entanglement between proton and neutron partitions for select $sd$-shell nuclides, for the empirical USDB interaction(solid lines) and for the isovector pairing Hamiltonian (dashed lines).  
    Other isotopic chains behavior in a qualitatively similar fashion.}
    \label{fig:entropy}
\end{figure}

\section{The weak entanglement approximation}

\label{WEA}

The nuclear shell-model basis states are typically written in bipartite fashion
by partitioning into proton and neutron components: $ | \alpha \rangle = | a
\rangle_\pi \otimes | i \rangle_\nu$. This in turn allows one to use ideas from
quantum information theory.  Specifically, the density matrix $\rho_{\alpha,
\beta} = c_\alpha c^*_\beta$ can also be written using these bipartite indices,
$\rho_{a i, b j} = c_{a i} c^*_{b j}$; then one can compute the reduced density
matrix by tracing over one of the partition indices:
\begin{equation}
\rho^\mathrm{red}_{a,b} = \sum_i \rho_{a i, b i} = \sum_i c_{a i} c^*_{b i}.
\end{equation} 
One can find the eigenvalues of the reduced density matrix, which is nothing
more than singular value decomposition (SVD), also called Schmidt
decomposition, and the SVD theorem tells us that it does not matter over which
partition index we trace.  While the trace of both $\rho$ and
$\rho^\mathrm{red} =1$, the eigenvalues of the former are 0 and 1, while the
eigenvalues $\lambda_r$ of the latter can be on the interval $(0,1)$. The
eigenspectrum can be characterized by the \textit{entanglement entropy},
\begin{equation}
S = - \sum_r \lambda_r \ln \lambda_r.
\end{equation}
$S=0$ means an unentangled system, one that can be written as a simple product
wave function. A system with a low $S$, relative to the maximum, we refer to as
``weakly entangled.'' This is not the same as weakly coupled; a system can be
strongly coupled yet weakly entangled. An example is a mean-field ansatz. 

Numerical experiments have shown that realistic shell-model wave functions have
low entropy, driven in part by shell structure~\cite{johnson2023proton};
indeed, compared to many other possible partitions of the basis space,
proton-neutron partitioning leads to the lowest
entropy~\cite{perez2023quantum}.  Furthermore, $N \neq Z$ systems have
significantly lower entropy than $N=Z$. This is good news, as heavier nuclides
which are more challenging to model are typically neutron-rich. 

Fig.~\ref{fig:entropy} shows the entropy of entanglement between the proton
and neutron partitions of the configuration-interaction wave functions in the
$sd$ valence space (with a frozen $^{16}$O core), for neon, sodium, and
magnesium isotopic chains. The wave functions were computed with the
high-quality empirical USDB interaction~\cite{PhysRevC.74.034315} as well as
with the schematic isovector pairing Hamiltonian. As one goes away from $N=Z$,
the left-most point of each line, the entropy decreases, often dramatically,
especially for the even-even nuclides, and less so for the odd-odd (sodium)
case. Other isotopic chains behave similarly.   Although not shown, the
attractive isoscalar quadrupole-quadrupole interaction does \textit{not}
result in similar behaviors.  These behaviors are seen empirically in other
valence spaces, such as the $pf$ shell, and even in cross-shell
spaces~\cite{johnson2023proton}.

\begin{figure}
    \centering
    \includegraphics[width=0.6\textwidth]{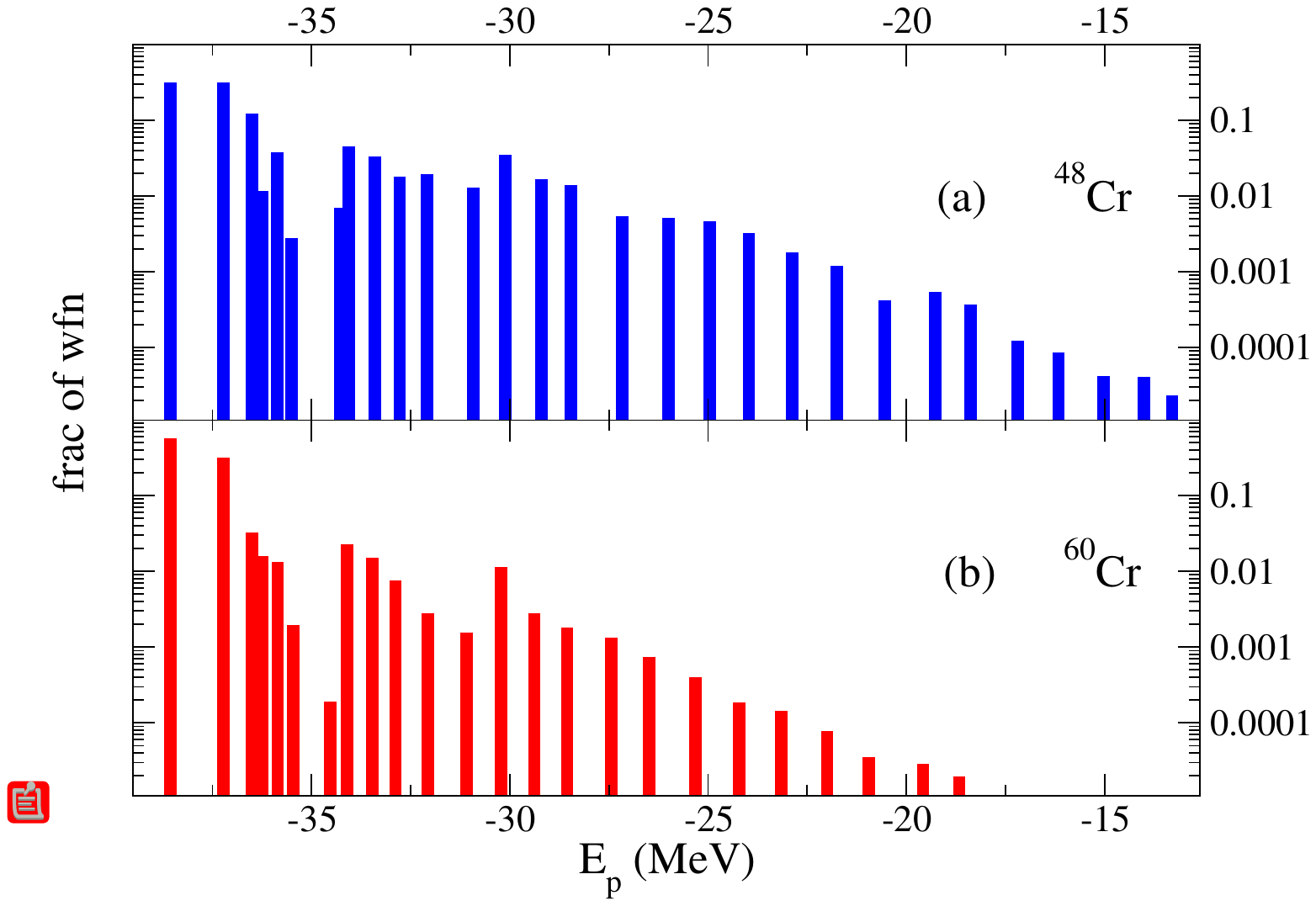}
    \caption{Decomposition of configuration-interaction wave functions for $^{48,60}$Cr: the fraction of the wave vector projected onto eigenstates of the many-proton components.  Note the the fast fall-off for 
    $^{60}$Cr, consistent with a lower entanglement entropy. }
    \label{fig:decomp}
\end{figure}

To exploit the  weak entanglement between the proton and neutron partitions
(see~\cite{PhysRevC.110.034305} for details), one  expands in a tensor product
basis:
\begin{equation}
| a \, J_a, i   \,  J_i: J \rangle = \left [ | a  \,  J_a \rangle_\pi \otimes | i   \, J_i \rangle_\nu \right ]_J, \label{pnbasis}
\end{equation}
where $| a   \,  J_a \rangle_\pi $ is a many-proton state with angular momentum
$J_a$ and label $a_\pi$, $| i  \,  J_i \rangle_\nu $ is a many-neutron state
with angular momentum $J_i$ and label $i_\nu$, coupled up to some total angular
momentum $J$.  We also couple parities but suppress that notation for
clarity. Working in such a  $J$-scheme basis, one expands 
\begin{equation}
| \Psi, J \rangle = \sum_{a,i} c_{a, i} |  a   \,  J_a, i  \,  J_i: J \rangle. \label{pnwfn}
\end{equation}
If one took all possible states $a, i$, we would recover the full configuration
interaction (FCI) space. (For comparison of current nuclear
configuration-interaction codes, the NuShellX code~\cite{NuShellX} works in
such a $J$-scheme, while the  $M$-scheme codes {\tt
BIGSTICK}~\cite{johnson2018bigstick}, KSHELL~\cite{shimizu2013nuclear}, and
ANTOINE~\cite{ANTOINE} codes all work with proton and neutron components in the
$M$-scheme.) 

Rather than taking all possible states, one can truncate  using only a select
set of the proton and neutron components.  This is not a new idea, but unlike
in some previous investigations  which iteratively optimized the
basis~\cite{papenbrock2003factorization,papenbrock2004solution,papenbrock2005density},
we  opt for a ``good enough'' basis.  One can justify this through a
straightforward numerical investigation.  Divide up the shell-model Hamiltonian
into proton, neutron, and proton-neutron sub-Hamiltonians, $\hat{H} = \hat{H}_p
+ \hat{H}_n + \hat{H}_{pn}$ (where $\hat{H}_p$ contains both one-body and
two-body contributions, and same for $H_{n}$; $\hat{H}_{pn}$ is 
only two-body). One can solve the proton and neutron Hamiltonians separately,
\begin{equation}
  \hat{H}_p | \phi_a, J_a \rangle_\pi 
  = E_a | \phi_a, J_a \rangle_\pi, \,\,\,\, \,\,\, \hat{H}_n | 
  \phi_i, J_i \rangle_\nu = E_i | 
  \phi_i, J_i \rangle_\nu; \label{pneigen}
\end{equation}
these proton and neutron eigenstates can be used to construct the basis as in
Eq.~(\ref{pnbasis}).  One  can  decompose the full proton-neutron wave vector,
Eq.~(\ref{pnwfn}),  and find the fraction associated with each proton (or
neutron)  eigenstate, that is, expressed as a function of the proton-sector
eigenenergy,
 \begin{equation}
f(a) = f(E_a) = \sum_i \left | c_{a,i} \right |^2.
\end{equation}
Even without explicit construction of this choice of basis, one can efficiently
carry out this decomposition using a version of the Lanczos
algorithm~\cite{PhysRevC.91.034313}. 

Fig.~\ref{fig:decomp}, decomposes the FCI wave vectors for $^{48}$Cr and
$^{60}$Cr computed in the $pf$ valence space using the $G$-matrix based
$pf$-shell  interaction GXPF1A~\cite{PhysRevC.65.061301,honma2005shell}.
$^{48}$Cr has four valence neutrons while $^{60}$Cr has four valence neutron
holes, meaning they have the same total dimension.  Overall one sees an
approximately exponential decrease in the component amplitudes, with a faster
decline associated with the $N > Z$ nuclide, along with a lower entropy. This
behavior is representative of a broader trend. 

\begin{figure}
  \begin{tabular}{cc}
    \includegraphics[width=0.52\textwidth]{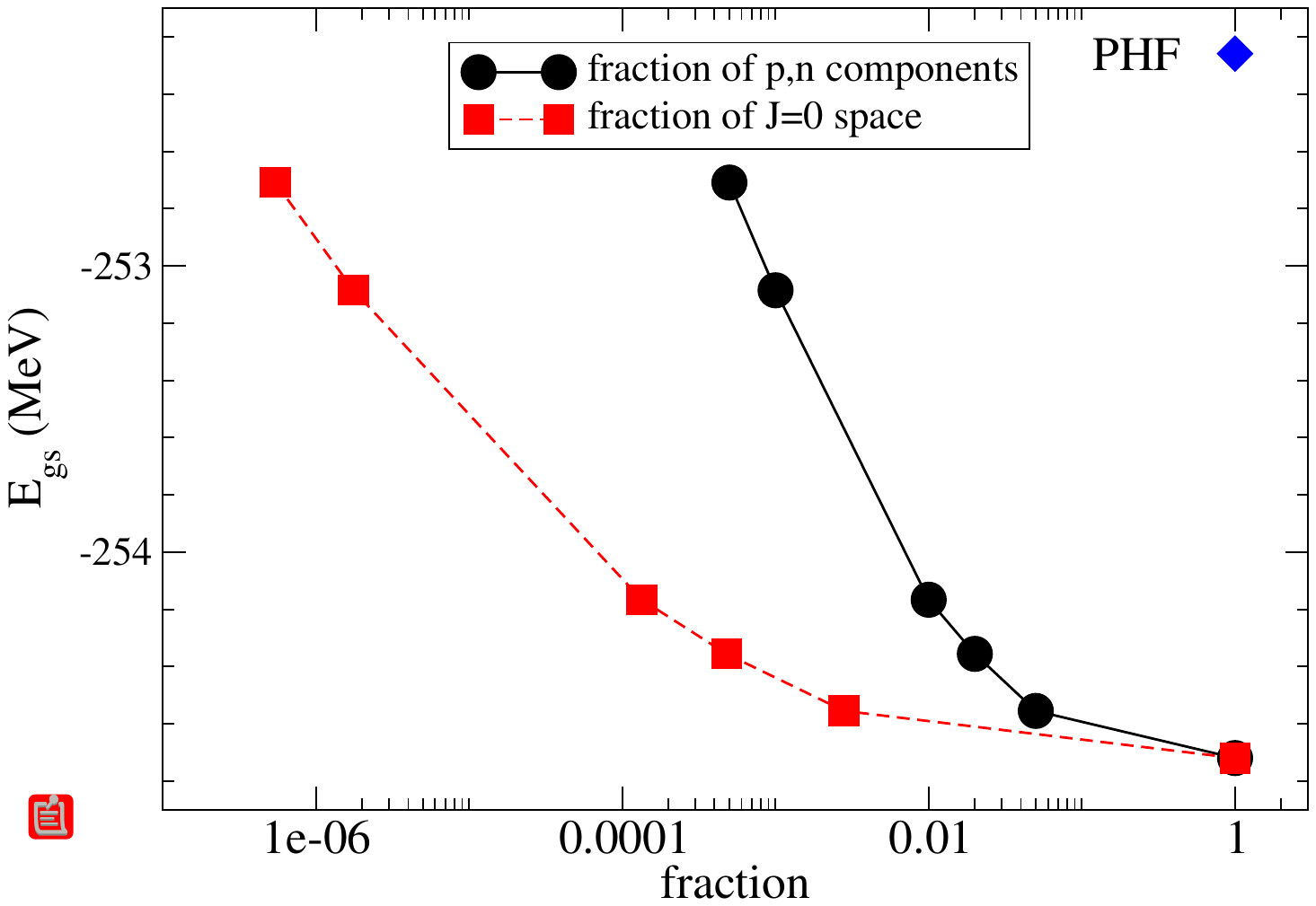} 
     \includegraphics[width=0.52\textwidth]{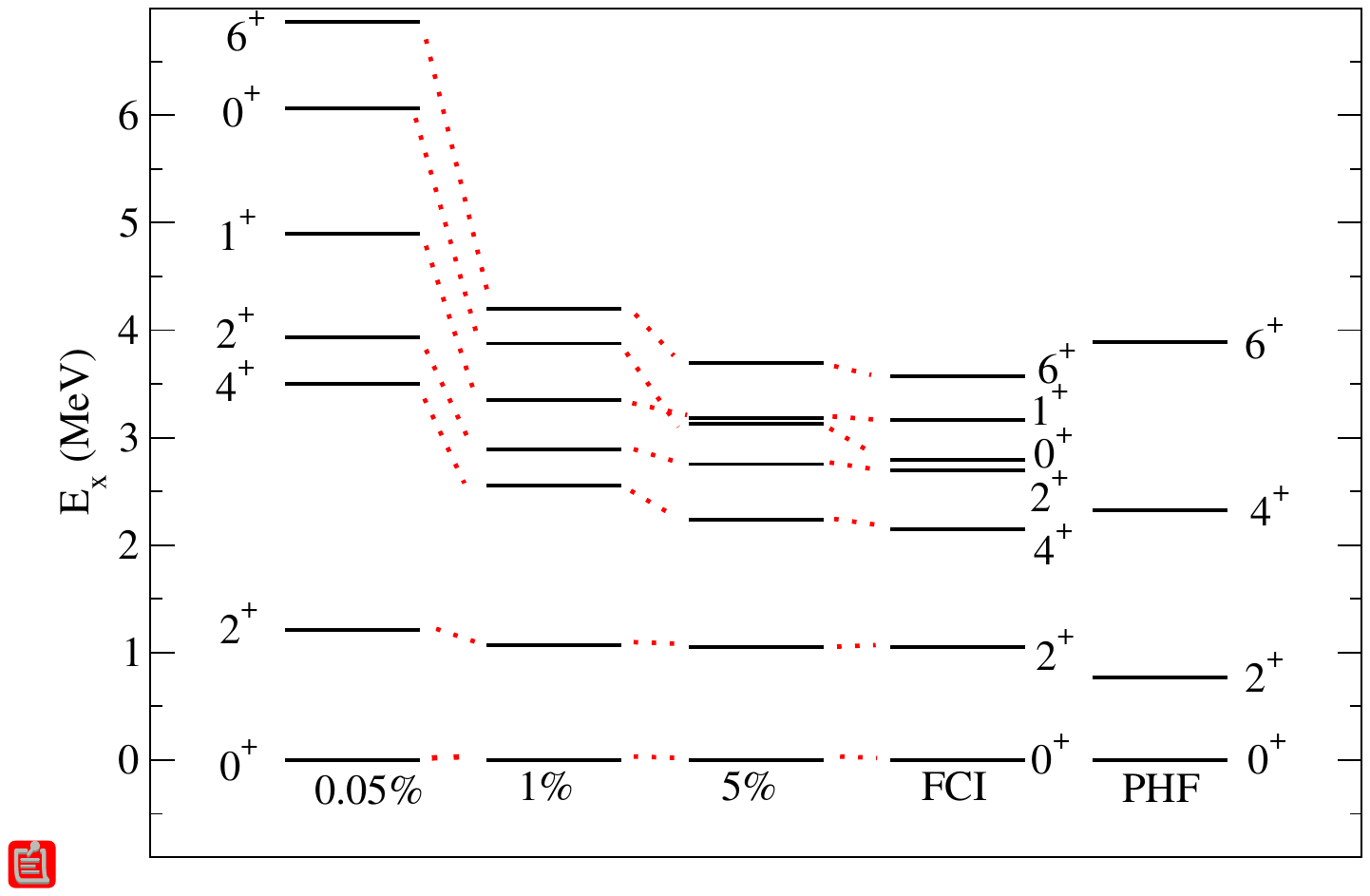}
  \end{tabular}
  \caption{Left: ground state energy of $^{60}$Zn in the $pf$ valence space,  as a function of the fraction of proton, neutron components used (black circles) and of the $J=0$ space (red squares). Here fraction = 1 is the full configuration-interaction result.  For comparison, we also give the angular-momentum projected Hartree-Fock (PHF) ground state energy (blue diamond).  Right: excitation energies of $^{60}$Zn computed in the $pf$ space. Shown are the approximate results using $0.05\%$ of the proton and neutron components, $1\%$, $5\%$, 
    the full configuration-interaction space (FCI), and angular-momentum projected Hartree-Fock (PHF).   }
  \label{fig:zn60dim}
\end{figure}

This exponential decay of component amplitudes leads to a practical methodology.  The Hamiltonian matrix is
block-diagonal in total angular momentum $J$ (and parity), with matrix elements 
\begin{eqnarray}
\langle a \, J_a, i \, J_i ; J | \hat{H} | b \, J_b , j \, J_j; J  \rangle &  = &   \delta_{i,j} \delta_{J_a, J_b} \langle a \, J_a  | \hat{H}_p | b \, J_b   \rangle +  \delta_{a,b} \delta_{J_i, J_j} \langle  i \, J_i  | \hat{H}_n |  j \, J_j  \rangle \nonumber \\
& & + \langle a \, J_a, i \, J_i ; J | \hat{H}_{pn} | b \, J_b , j \, J_j; J  \rangle; \label{Hpanash}
\end{eqnarray}
the key proton-neutron matrix element can be expressed  in terms of one-body
transition density matrices; see ~\cite{PhysRevC.110.034305} for details.  

If one chooses the states $ |  a \, J_a \rangle_\pi$ to be eigenstates of
$\hat{H}_p$, and similarly the states $| i \, J_i \rangle_\nu$ to be eigenstates
of $\hat{H}_n$, results such as Fig.~\ref{fig:decomp} justify truncating on the
basis of the energies of the proton and neutron components. Furthermore, in such
a case the matrix elements of Eq.~(\ref{Hpanash}) are further simplified.  The
required eigenenergies can be produced as a matter of course in an $M$-scheme
code such as {\tt BIGSTICK}, and the one-body transitions densities can also be
produced routinely. The dimensions, however, are far smaller, as discussed
below.  The main challenge, discussed in the Appendix, is being able to generate
a sufficient number of basis components.

Initial work has shown this truncation scheme provides a good approximation for
energies. Nonetheless, one can use more sophisticated choices for the proton and
neutron basis states, an area of very near-future exploration.

\section{Results and conclusion}
\label{results}

As a first example, consider $^{60}$Zn which has 10 protons and 10 neutrons in
the $pf$ valence space. In the $M$-scheme, the basis dimension for $^{60}$Zn
with $M=0$ is 2.2 billion, the largest in the space. However the basis dimension
for 10 protons in the  $pf$ shell is, for $M =0$, only 17,276, and the same for
10 neutrons.  Fig.~\ref{fig:zn60dim} shows the computed ground state energy,
using the GXPF1A interaction, as a function of the fraction of the basis proton
and neutron components, starting with only nine many-proton and nine
many-neutron states.  The ground state energies are also given as a fraction of
full configuration-interaction (FCI) $J=0$ dimension (31 million). The
$J$-scheme dimensional scales approximately as the product of the number of
proton and the number of neutron components.  For comparison, we also give the
ground state energy computed using unrestricted (no assumption of axial or other
symmetries) angular-momentum projected (after variation) Hartree-Fock (PHF)
using the same shell-model inputs~\cite{lauber2021benchmarking}.  Even the small
PANASh case building a basis from nine proton components and nine neutron
components outperforms PHF.

Excitation energies of $^{60}$Zn are also shown in Fig.~\ref{fig:zn60dim} for
select fractions of the proton and neutron components, as well as the PHF
excitation spectrum.  The smallest PANASh calculation, using  $0.05\%$ of the
proton and neutron components, reproduces the qualitative features of the
excitation spectrum; at $1\%$ of the components one sees a very good
reproduction of the excitation spectrum.

\begin{figure}
    \begin{tabular}{cc}   
      \includegraphics[width=0.5\textwidth]{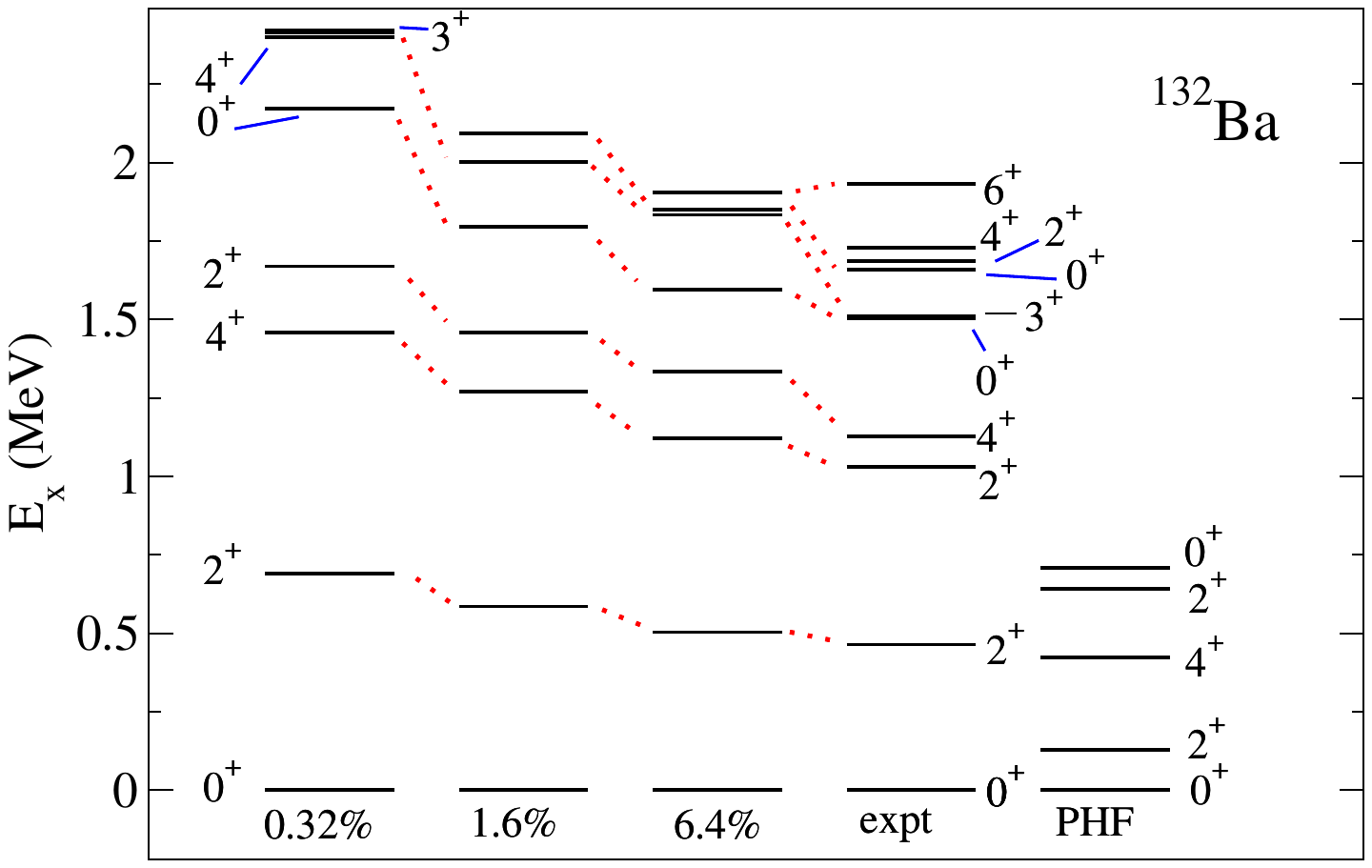} 
      \includegraphics[width=0.5\textwidth]{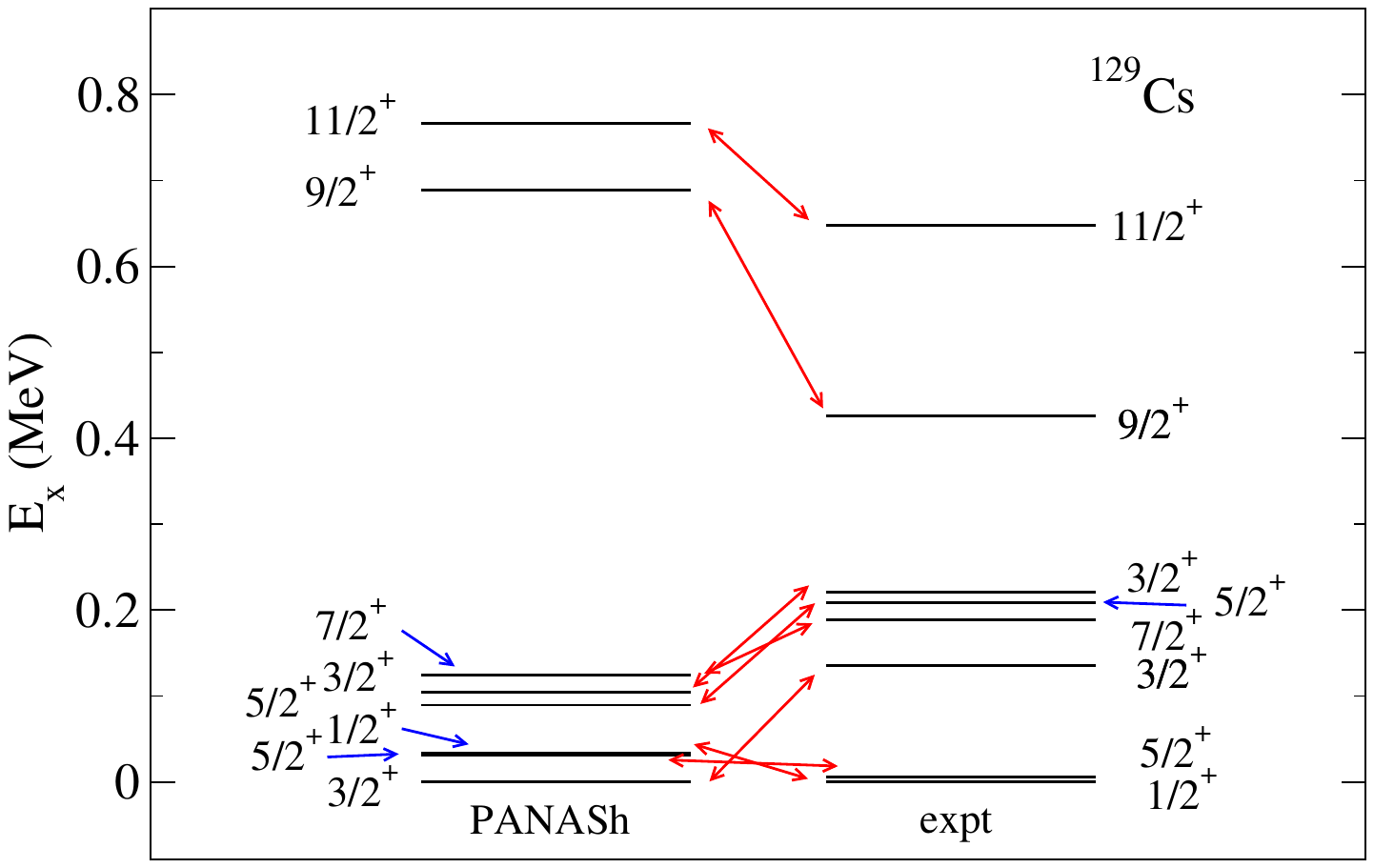}
    \end{tabular}
    \caption{Left: Excitation energies of $^{132}$Ba. Right: $^{129}$Cs.  Both are computed assuming a $^{100}$Sn core and valence orbitals between shell closures at 50 and 82, using the GCN5082 interaction~\cite{Caurier:2007wq,Caurier:2010az}.
    The FCI dimensions are 20 billion and 50 billion, respectively.  For comparison, we also show the angular-momentum projected Hartree-Fock (PHF) excitation spectrum for $^{132}$Ba; the PHF spectrum for $^{129}$Cs, 
    not shown, is even worse. }
    \label{fig:shell5082}
\end{figure}

Fig.~\ref{fig:shell5082} shows excitation spectra for two nuclides, $^{132}$Ba
and $^{129}$Cs.  These are computed with a $^{100}$Sn core and a valence space
of $0g_{7/2}$-$2s_{1/2}$-$1d_{3/2,5/2}$-$0h_{11/2}$, using the GCN5082
interaction~\cite{Caurier:2007wq,Caurier:2010az}. The FCI dimensions are 20
billion and 50 billion;  the latter is just beyond current computational
resources. Therefore we compare not to FCI results but to experiment. 

The PANASh spectrum of $^{132}$Ba is  good. Even the smallest fraction
reproduces the level ordering, and with $1.6\%$ of the proton and neutron
components, one gets a very good approximation. (For comparison, we also include
the PHF excitation spectrum. The PHF result has much poorer agreement, which we
attribute to lack of pairing correlations~\cite{lauber2021benchmarking}.) 

For $^{129}$Cs, we used 1000 proton components ($6.8\%$) and 1000 neutron
components ($0.15\%$). (Preliminary results suggests a cutoff based upon
energy,  not fractions, yields better results. This will be a focus of future
work.) As expected for  odd-$A$, the density of levels is much higher, so it is
not surprising that we do not reproduce the spectrum with high accuracy.
Nonetheless we reproduce approximately the ordering of levels and spacings. 
Although not shown, a PHF excitation spectrum is significantly more compressed. 
 
To summarize, we have provided motivation for and implementation of a `weak
entanglement' approximation to the configuration-interaction shell model, and
provide examples not previously published \cite{PhysRevC.110.034305}.

\section{Acknowledgements}

This work was performed in part under the auspices of the U.S. Department of
Energy by Lawrence Livermore National Laboratory under Contract
DE-AC52-07NA27344 with support from the Weapon Physics and Design (WPD) Academic
Collaboration Team (ACT) University Collaboration program.  This material is
also based upon work supported by the U.S. Department of Energy, Office of
Science, Office of Nuclear Physics, under Award Number DE-FG02-03ER41272.  Part
of this research was enabled by computational resources supported by a gift to
SDSU from John Oldham.  CWJ thanks Ken McElvain for a  useful conversation
regarding  LAPACK eigensolvers. 

\appendix

\section{Some technical issues}

\label{technical}

Here we discuss some technical issues and how we recently overcame them. 

Although PANASh is an efficient approximation, it nonetheless  requires a large
number of input basis levels, as well as the one-body density matrices between
them.  If the model space allows for  positive and negative parities, one
needs levels of both parities and inter- and intra-parities densities.  To
generate the basis levels, we use the {\tt BIGSTICK} configuration-interaction
code~\cite{BIGSTICK}. It is an $M$-scheme code, which means the basis has fixed
total $M$, the $z$-component of angular momentum.  One can also fix the parity
or allow for both parities, but the latter option doubles the basis dimensions.
{\tt BIGSTICK} can compute one-body densities, but with the restriction that
all states must be in the same basis. 

{\tt BIGSTICK} produces reduced density matrices, which, according to the Wigner-Eckart
theorem~\cite{edmonds1996angular}, are independent of $M$. Nonetheless, the densities are computed at a fixed $M$, dividing by a Clebsch-Gordan coefficient 
to remove the dependence on orientation. Some coefficients, however, must vanish when $M=0$, which in turns can lead
to missing density matrices.  One option is to run both $M=0$ and $M=1$, but not
only is this inefficient, for large numbers (1000s) of levels, small differences
in convergence can make matching levels problematic.

To address these issues, we use an unpublished post-processing code, {\tt
RHODIUM}. Unlike {\tt BIGSTICK}, {\tt RHODIUM} can compute densities between
states in different bases.  By using an angular momentum raising operator, i.e.,
$\hat{J}_+$, we can generate from $M=0$ wave vectors the corresponding $M=1$
wave vectors and regain the missing density matrices without rerunning {\tt
BIGSTICK}. This also eliminates the issue of matching levels between two
different large-scale runs.  {\tt RHODIUM} also allows us to directly compute
density matrices between wave vectors computed in basis with opposite parities,
another saving.

Another issue is the time-to-solution for generating the base levels. If one
wants to generate, say, 1000 converged levels, one needs  $\sim 5000$ or more
Lanczos iterations. {\tt BIGSTICK} checks converges by comparing  the first
$N_\mathrm{keep}$  eigenvalues, where $N_\mathrm{keep}$ is the number of desired
converged states.  This can add significantly to the run time.  We found an
efficient solution: rather than checking the convergence of all
$N_\mathrm{keep}$ levels, we instead track the convergence of the last
$N_\mathrm{test}$ of them, that is, the convergence of eigenvalues
$N_\mathrm{keep} - N_\mathrm{test}+1$ to $N_\mathrm{keep}$. This can be done
very efficiently on Lanczos tridiagonal matrices using the LAPACK routine {\tt
DSTEGR}; we found choosing $N_\mathrm{test} = \sqrt{ N_\mathrm{keep}}$ worked
well. Furthermore, when extracting the final eigenvectors, the LAPACK routine
{\tt DSYEVR} is more efficient for generating the first $N_\mathrm{keep}$
eigenvectors.  These technical improvements will enable us to generate PANASh
solutions more efficiently and to achieve larger cases with the same
computational resources.

\bibliographystyle{elsarticle-num}
\bibliography{johnsonmaster}

\end{document}